\documentclass[12pt,draftclsnofoot,peerreviewca,onecolumn,letterpaper]{IEEEtran}

\usepackage{amsfonts,amsmath,amssymb,bbm}
\usepackage{graphicx,color,epsfig,rotating}
\usepackage{latexsym}
\usepackage{subfigure}
\usepackage{cite}
\usepackage{epic,eepic}

\long\def\comment#1{}

\newfont{\bbb}{msbm10 scaled 700}

\newfont{\bb}{msbm10 scaled 1100}
\newcommand{\CC}{\mbox{\bb C}}
\newcommand{\PP}{\mbox{\bb P}}
\newcommand{\RR}{\mbox{\bb R}}

\newcommand{\ZZ}{\mbox{\bb Z}}

\newcommand{\EE}{\mbox{\bb E}}

\newcommand{\hv}{{\bf h}}

\newcommand{\rv}{{\bf r}}

\newcommand{\vv}{{\bf v}}
\newcommand{\xv}{{\bf x}}

\newcommand{\zerov}{{\bf 0}}
\newcommand{\onev}{{\bf 1}}

\newcommand{\Am}{{\bf A}}

\newcommand{\Hm}{{\bf H}}
\newcommand{\Id}{{\bf I}}

\newcommand{\Qm}{{\bf Q}}
\newcommand{\Rm}{{\bf R}}

\newcommand{\Ac}{{\cal A}}

\newcommand{\Cc}{{\cal C}}

\newcommand{\Nc}{{\cal N}}

\newcommand{\Rc}{{\cal R}}
\newcommand{\Sc}{{\cal S}}

\newcommand{\gammav}{\hbox{\boldmath$\gamma$}}

\newcommand{\Sigmam}{\hbox{\boldmath$\Sigma$}}

\newcommand{\trace}{{\hbox{tr}}}

\renewcommand{\arg}{{\hbox{arg}}}

\newcommand{\SINR}{{\sf sinr}}

\newcommand{\eqdef}{\stackrel{\Delta}{=}}

\newcommand{\herm}{{\sf H}}

\newtheorem{thm}{Theorem}
\newtheorem{lemma}{Lemma}

\def\BibTeX{{\rm B\kern-.05em{\sc i\kern-.025em b}\kern-.08em
    T\kern-.1667em\lower.7ex\hbox{E}\kern-.125emX}}

\begin{document}

\title{Joint Scheduling and ARQ for MU-MIMO Downlink in the Presence of Inter-Cell Interference}
\author{\IEEEauthorblockN{H.Shirani-Mehr\IEEEauthorrefmark{1}, H. Papadopoulos\IEEEauthorrefmark{2},
S. A. Ramprashad\IEEEauthorrefmark{2}, G. Caire\IEEEauthorrefmark{1}}
\IEEEauthorblockA{\IEEEauthorrefmark{1}University of Southern California, Email: shiranim, caire@usc.edu}
\IEEEauthorblockA{\IEEEauthorrefmark{2}DoCoMo Laboratories USA, Inc., Email: ramprashad, hpapadopoulos@docomolabs-usa.com}
}
\thispagestyle{empty}

\maketitle

\begin{abstract}
User scheduling and multiuser multi-antenna (MU-MIMO) transmission are
at the core of high-rate data-oriented downlink schemes
of the next-generation of cellular systems (e.g., LTE-Advanced).
Scheduling selects groups of users according to their channels vector directions
and SINR levels. However, when scheduling is applied independently in each cell, the inter-cell interference (ICI) power
at each user receiver is not known in advance since it changes at each new scheduling slot
depending on the scheduling decisions of all interfering  base stations.
In order to cope with this uncertainty, we consider the joint operation of scheduling, MU-MIMO beamforming and
Automatic Repeat reQuest (ARQ).  We develop a game-theoretic framework for this problem
and build on stochastic optimization techniques in order to find optimal scheduling and ARQ schemes.
Particularizing our framework to the case of ``outage service rates'', we obtain
a scheme based on adaptive variable-rate coding at the physical layer, combined with ARQ at the Logical Link Control (ARQ-LLC).
Then, we present a novel scheme based on incremental redundancy Hybrid ARQ (HARQ) that is able to achieve a
throughput performance arbitrarily close to the ``genie-aided service rates'', with no need for a genie that provides
non-causally the ICI power levels. The novel HARQ scheme is both easier to implement
and superior in performance with respect to the conventional combination of adaptive variable-rate
coding and ARQ-LLC.
\end{abstract}


\newpage

\section{Introduction} \label{intro}

High-rate data-oriented downlink schemes \cite{HDR,CDMA-HDR-Jalali}
have been successfully deployed as an extension of 3G cellular standards
(WCDMA and CDMA2000).
These schemes are based on the results of \cite{Tse-Unpublished,Tse-Hanly,tse-opportunistic}, showing that
the throughput (or ``ergodic'')  sum-capacity  of single-antenna multi-access
(uplink) and  broadcast (downlink) fading Gaussian channels is achieved
by allocating opportunistically each time-frequency slot to the user with the best instantaneous
channel conditions.  In a multiuser setting, the sum-capacity is usually not the most meaningful
measure of the system performance. Instead,  maximizing the sum-throughput subject to
some {\em fairness} constraint is more desirable \cite{tse-opportunistic}. To this purpose, a {\em downlink scheduling policy}
can be designed  in order to maximize a suitable concave and component-wise monotonically
increasing network utility function
over
the system's achievable throughput region (i.e., the region of  achievable long-term average user rates).
The network utility function is designed in order to capture the desired notion of ``fairness'' (e.g.,
proportional fairness, max-min fairness and, more in general, $\alpha$-fairness \cite{Mo-Walrand}).

In the next generation of cellular systems (e.g., the so-called LTE-Advanced \cite{LTE-advanced}),
high-rate data-oriented downlink schemes will be combined with multiuser multi-antenna
(MU-MIMO) transmission techniques \cite{caire2003atm,weingarten2004crg},
supporting spectral efficiencies in the 10's of bits/sec/Hz  \cite{ramp-caire-PIMRC09,Foschini-Karakayali-Valenzuela-TIEEProcComm06}.
With MU-MIMO, the rate supported by each user is generally a function of {\em all} the user channel vectors,
and depends on the type of MU-MIMO precoding  \cite{caire2003atm, weingarten2004crg,tosato-ita}.
In order to compute the transmitter precoder parameters (e.g., the beamforming steering
vectors and the transmitted rates and powers), channel state information at the transmitter (CSIT) is required.
This can be accurately obtained using open and closed loop channel estimation and feedback schemes (the literature on this subject is
overwhelming, for example, see \cite{marzetta2006mtr,Caire-Jindal-Kobayashi-Ravindran,ding2007mab,Kobayashi-Caire-Training,HoomanTcom}
and references therein).

In particular, scheduling with MU-MIMO and non-perfect CSIT
was considered in \cite{Hooman-nonperfectCSI}, particularizing the general stochastic optimization framework of
\cite{NeelyBook} to the case of a single-cell system with linear Zero-Forcing Beamforming (ZFBF) MU-MIMO precoding,
where CSIT is obtained via noisy channel estimation and prediction.

In this work we focus on a multi-cell environment with no inter-cell cooperation.
For sufficiently slowly-moving user terminals it is possible to design training and feedback schemes that achieve
almost  perfect CSIT
\cite{Kobayashi-Caire-Training,HoomanTcom,Hooman-nonperfectCSI}.
Therefore, for simplicity we shall assume that each BS has perfect CSIT for its own users.
In contrast, in a multi-cell system, inter-cell interference (ICI) emerges as another source of unavoidable uncertainty.
(see \cite{Fodor-Koutsimanis,DBLP:conf/globecom/IvrlacN07} and references therein).
When the schedulers at each BS make their own decisions independently,
based only on the locally available CSIT relative to their own users,
the ICI power seen at each user receiver changes on a slot-by slot basis
in a random and unpredictable manner, depending on the scheduling decision made at all the interfering BSs.
As a consequence, the instantaneous Signal to Interference plus Noise Ratio (SINR)
``seen'' at any given user receiver is a random variable.

The decentralized scheduling problem in a multi-cell environment
can be formulated as a non-cooperative game: each BS (player) wishes to maximize its own utility function
over its own feasible throughput region.  The players' strategies are all feasible scheduling policies.
In addition, the throughput region of any given cell depends on the ICI power statistics seen at the users' receivers,
which in turn depend on the scheduling policies applied at the interfering BSs.
We show that when the individual network utility functions are concave
the multi-cell decentralized scheduling game is a concave game and therefore
Nash equilibria exist.

In order to solve the network utility maximization at each BS, for given ICI statistics,
we apply the stochastic optimization framework of \cite{NeelyBook, neely-fairness-infocom05,NeelyJSAC05,Hooman-nonperfectCSI}.
A straightforward application of this approach yields a scheme based on variable-rate adaptive coding
at the physical layer, and conventional ARQ at the Logical Link Control (LLC) layer.
We notice that similar approaches are included in several wireless standards such as EV-DO and
HSDPA \cite{EVDO-HDR,HARQ-CC-HSDPA,3GPP-HSDPA}, and therefore this can be regarded as
the base-line ``conventional'' approach.  In order to improve upon the conventional approach,
we propose a new method based on combining incremental redundancy Hybrid  Automatic Retransmission reQuest (HARQ) \cite{caire-tuninetti} and
MU-MIMO opportunistic scheduling. In the proposed scheme, each user feeds back the value of the instantaneous
mutual information observed in the previous slot, that is used by the scheduler to update recursively
the scheduler weights. We show that the throughput achieved by the proposed HARQ scheme
approaches the throughput of a ``virtual system'', as if  a genie provided non-causally the ICI values at each scheduling slot.
However, we stress that the proposed scheme makes use of strictly causal information, and therefore
requires no genie.


\section{System setup} \label{model}

We consider the downlink of a system with $C > 1$ cells. In each cell, a BS equipped with
with $M$ antennas transmits to $K$ single-antenna users.
The channel is assumed frequency flat\footnote{The generalization to MIMO-OFDM and frequency selective fading is immediate.}
and constant over ``slots'' of length $T \gg 1$ symbols (block-fading model \cite{biglieri1998fci}).
Any given channel use of the complex baseband discrete-time signal at the $k$-th user in cell $c$ during slot $t$ is described by
\begin{equation} \label{mimo-ofdm}
y_{k,c}[t] = \underbrace{\sqrt{g_{k,c,c}} \hv^\herm_{k,c,c}[t] \xv_c[t]}_{\mbox{desired BS}} + \underbrace{\sum_{c' \neq c} \sqrt{g_{k,c,c'}}
\hv^\herm_{k,c,c'}[t] \xv_{c'}[t]}_{\mbox{inter-cell interference}}  + z_{k,c}[t],
\end{equation}
where $t$ ticks at the slot rate,  $(k,c)$ denotes user $k$ in cell $c$,
$\hv_{k,c,c'}[t] \in \CC^M$ is the channel vector from the $c'$-th
BS antenna array to the $(k,c)$-th receiver antenna,
$\xv_{c'}[t] \in \CC^M$ is the signal
transmitted by $c'$-th BS and $z_{k,c}[t] \sim \Cc\Nc(0,1)$ is the additive white Gaussian noise (AWGN) sample.
The coefficients $g_{k,c,c'}$ are distance-dependent path gains \cite{goldsmith2005wc} that are assumed to be time-invariant
over many slots. The BSs are sum-power constrained  such that $\trace \left ( \Sigmam_c[t] \right ) \leq 1$ for all $t$,
where $\Sigmam_c[t] =  \EE[ \xv_c[t] \xv^\herm_c[t]]$ denotes the transmit covariance matrix.
The actual channel SNR is included as a common scaling factor in the coefficients $g_{k,c,c'}$.
The channel vectors of users in cell $c$ form the columns of  the channel matrix
$\Hm_c [t] = [\hv_{1,c,c} [t],...,\hv_{K,c,c} [t]] \in \mathbb{C}^{ M \times K }$.
We assume that all vectors $\hv_{k,c,c'}[t]$ are mutually independent with
i.i.d. components $\sim \Cc\Nc(0,1)$, for all distinct 4-tuples $(t,k,c,c')$.
Each BS $c$ knows all time-invariant quantities relative to its own users
and has perfect knowledge of $\Hm_c[t]$ immediately before the beginning of slot $t$ (perfect CSIT for the own users).

A feasible scheduling policy $\gamma_c$ for BS $c$ is a possibly {\em randomized} stationary
function\footnote{Using the theory developed in \cite{DBLP:journals/tit/Neely06}
we can show that restricting to {\em stationary} policies does involve any suboptimality in terms of the achievable throughput region.}
that maps $\Hm_c[t]$  into the pair $\gamma_c(\Hm_c[t]) = (\Sigmam_c[t], \rv_c[t])$, where
$\rv_c[t] = (r_{1,c}[t], \ldots, r_{K,c}[t])$ is a rate allocation vector.
We assume that the MU-MIMO precoder is based on linear ZFBF. This yields the transmitted signal vector in the form
$\xv_c[t] = \sum_{k \in \Sc_c[t]}  \vv_{k,c}[t] u_{k,c}[t]$, where $\Sc_c[t]$ denotes the set of active users, i.e., users that are
selected to be served on slot $t$ and where $u_{k,c}[t] \in \CC$ denotes the coded symbol for user $(k,c)$, with power
$\EE[|u_{k,c}[t]|^2] = P_{k,c}[t]$. The ZFBF steering vectors $\{\vv_{k,c}[t]: k \in \Sc_c[t]\}$ are given by the
unit-norm (normalized) $k$-th column of the Moore-Penrose pseudoinverse (e.g., see \cite{Caire-Jindal-Kobayashi-Ravindran,jose:csi,ding07,Kobayashi-Caire-Training,Hooman-nonperfectCSI}  and references therein) of the channel matrix restricted to the active users, i.e., to the columns $\{\hv_{k,c,c}[t] : k \in \Sc_c[t]\}$. It follows that  the transmit covariance matrix takes on the form
\begin{equation} \label{tx-cov}
\Sigmam_{c}[t] = \sum_{k \in \Sc_c[t]} \vv_{k,c}[t] \vv_{k,c}^\herm[t] P_{k,c}[t].
\end{equation}
where non-negative coefficients $\{P_{k,c}[t] : k \in \Sc_c[t]\}$ define the power allocation over
the active users in cell $c$, and satisfy the power constraint $\sum_{k \in \Sc_c[t]} P_{k,c}[t] \leq 1$.
A necessary and sufficient condition for perfect zero-forcing of the intra-cell multiuser interference
is that $|\Sc_c[t]| \leq \min\{M,K\}$.  Without loss of generality, in the following we
identify the set of active users $\Sc_c[t]$ with those users with positive powers, i.e., $P_{k,c}[t] > 0$ for $k \in \Sc_c[t]$
and $P_{k,c}[t] = 0$ for $k \notin \Sc_c[t]$.

The ICI power at user $(k,c)$ receiver in slot $t$ is given by
\begin{equation} \label{inst-ici}
\chi_{k,c}[t] = \sum_{c' \neq c} g_{k,c,c'}  \hv_{k,c,c'}^\herm[t] \Sigmam_{c'}[t] \hv_{k,c,c'}[t]
\end{equation}
with mean given by
$\overline{\chi}_{k,c} = \EE[ \chi_{k,c}[t]] = \sum_{c' \neq c} g_{k,c,c'} \trace(\Sigmam_{c'}[t])$.
The SINR at user $(k,c)$ is given by
\begin{eqnarray}\label{SINR}
\SINR_{k,c}[t] & = &  \frac{g_{k,c,c} \left |\hv_{k,c,c}^\herm[t] \vv_{k,c}[t] \right |^2 P_{k,c}[t] }{1 + \chi_{k,c}[t]}
\end{eqnarray}
We let $R_{k,c}[t]$ denote the {\em instantaneous service rate} of user $(k,c)$ on slot $t$, measured in bits/channel use.
This is in general a function of $\SINR_{k,c}[t]$, and therefore of $\Hm_c[t], \Sigmam_c[t], \chi_{k,c}[t]$, and of
the allocated rate  $r_{k,c}[t]$.  We define the $k$-th user service rate function $R_k(g,\Hm, \chi, \Sigmam,\rv)$, such that
$R_{k,c}[t] = R_k(g_{k,c,c}, \Hm_c[t], \chi_{k,c}[t], \Sigmam_c[t], \rv_c[t])$.
Let $\Gamma$ denote the set of all feasible scheduling policies and let
$\gamma_{-c} = \{\gamma_{c'} : c' \neq c\}$ denote the set of scheduling policies at all cells
$c' \neq c$. For fixed  $\gamma_{-c} \in \Gamma^{C-1}$,  the throughput of user $(k,c)$ under the scheduling policy $\gamma_c$  is given by
\begin{eqnarray} \label{throughput}
\overline{R}_{k,c}(\gamma_c, \gamma_{-c}) & = & \liminf_{t \rightarrow \infty} \frac{1}{t} \sum_{\tau=1}^t R_k(g_{k,c,c}, \Hm_c[\tau], \chi_{k,c}[\tau], \gamma_c(\Hm_c[\tau])) \nonumber \\
& = & \EE \left [ R_k(g_{k,c,c}, \Hm_c, \chi_{k,c}, \gamma_c(\Hm_c)) \right ]
\end{eqnarray}
where the effect of the policies at the interfering BSs is captured by the statistics of the ICI power process
$\chi_{k,c}[t]$, the limit holds almost surely because of stationarity and ergodicity, and
expectation is with respect to the joint distribution of the triple $(\Hm_c[t], \chi_{k,c}[t],\gamma_c)$.
\footnote{With some abuse of notation, we denote by  $\Hm_c$ and $\{\chi_{k,c} : k = 1, \ldots, K\}$ random variables whose joint distribution
coincides with the first-order joint distribution of the processes $\Hm_c[t]$ and $\{\chi_{k,c}[t] : k = 1, \ldots, K\}$, which is time-invariant
by stationarity.}
The region of achievable throughputs for cell $c$ is given by
\begin{equation} \label{ergodic-region}
\Rc_c(\gamma_{-c}) = {\rm coh} \bigcup_{\gamma_c \in \Gamma} \left \{ \overline{\Rm} \in \RR_+^K :
\overline{R}_k \leq \EE \left [R_k(g_{k,c,c},\Hm_c, \chi_{k,c}, \gamma_c(\Hm_c)) \right ] , \;\; \forall \; k \right \}
\end{equation}
``coh'' denotes ``closure of the convex hull''. Notice that $\Rc_c(\gamma_{-c})$ depends on the other cells' scheduling policies $\gamma_{-c}$
through the  joint probability distribution of the ICI powers $\{\chi_{k,c} : k = 1, \ldots, K\})$.

Under our assumptions, the BSs operate in a decentralized way and influence each other only in terms of
the generated ICI statistics (i.e., the joint cdfs $\{\chi_{k,c} : k = 1, \ldots, K\}$).
Each BS wishes to maximize its own network utility function. This multi-objective optimization problem
is formulated as a non-cooperative game \cite{friedman-71,Friedman1977} that we nickname the {\em multi-cell decentralized scheduling game},
where each player (i.e., BS) $c$ seeks to achieve
\begin{eqnarray} \label{num-max1}
\mbox{maximize} & & U_c(\overline{\Rm}) \nonumber \\
\mbox{subject to} & & \overline{\Rm} \in \Rc_c(\gamma_{-c})
\end{eqnarray}
where we assume that $U_c(\cdot)$ is a continuous, strictly concave and component-wise increasing utility function,
reflecting some suitable fairness criterion \cite{Mo-Walrand}.

By definition, for any given joint statistics of $\Hm_c$ and of $\{\chi_{k,c} : k = 1,\ldots, K\}$,
the maximum in (\ref{num-max1}) is achieved by  some scheduling policy $\gamma_c^\star$.
A Nash equilibrium of the {\em decentralized scheduling game} is a set of scheduling policies (also denoted, with some abuse of notation,
by $\{\gamma_c^\star : c = 1, \ldots, C\}$)  such that $\gamma_c^\star$ is the solution to (\ref{num-max1})
when $\gamma_{-c} = \gamma_{-c}^\star$,  for all $c = 1,\ldots, C$. We have:

\begin{thm} \label{nash-existence}
The decentralized scheduling game defined above is a concave game and therefore has a Nash equilibrium.
\end{thm}

\begin{IEEEproof}
All players  have the same strategy set $\Gamma$. This is a compact convex set due to the
covariance trace constraint and to the fact that we can assume that the rate allocation vector
is bounded in $\rv_c \in [0, r_{\max}]^K$  for some constant $r_{\max}$.
\footnote{This limitation does not involve any significant  loss
of generality if $r_{\max}$ is sufficiently large, and always holds in practice since practical variable-rate coding
has a finite maximum rate.}
Also, each $c$-th utility is a concave function of $\gamma_c$ for fixed $\gamma_{-c}$.
In order to see this,  let $\overline{\Rm}(\gamma_c,\gamma_{-c})$ denote the throughput point
of $\Rc_c(\gamma_{-c})$ achieved by policy $\gamma_c$ for fixed $\gamma_{-c}$,
consider any two policies $\gamma'_c, \gamma''_c \in \Gamma$ and define $\gamma^{(\lambda)}_c$ as the policy that
applies $\gamma'_c$ with probability $\lambda \in [0,1]$ and $\gamma''_c$ with probability $\bar{\lambda} = 1 - \lambda$.
Then, from the convexity of $\Rc_c(\gamma_{-c})$ and the concavity of $U_c(\cdot)$ we have that
\[ \lambda U_c(\overline{\Rm}(\gamma'_c,\gamma_{-c})) + \bar{\lambda} U_c(\overline{\Rm}(\gamma''_c,\gamma_{-c}))
\leq U_c( \lambda \overline{\Rm}(\gamma'_c,\gamma_{-c}) + \bar{\lambda} \overline{\Rm}(\gamma''_c,\gamma_{-c}))
= U_c(\overline{\Rm}(\gamma^{(\lambda)}_c,\gamma_{-c})) \]
Now, let $\gammav = \{ \gamma_c : c = 1,\ldots, C\}$ and
$\gammav' = \{ \gamma'_c : c = 1,\ldots, C\}$ denote two vectors of scheduling policies and define the sum-utility function
$\rho(\gammav, \gammav') = \sum_{c=1}^C U_c(\overline{\Rm}(\gamma_c, \gamma'_{-c}))$.
Since the functions $U_c(\cdot)$ are continuous (by assumption) and the throughput vectors are
continuous functions of the scheduling policies, it follows that $\rho(\gammav, \gammav')$ is a continuous function of
$(\gammav, \gammav') \in \Gamma^C \times \Gamma^C$ and, for what said before,
it is concave in $\gammav$ for any fixed $\gammav'$. These properties match exactly the assumption of
Rosen Theorem \cite{Rosen1965}. Therefore, as a direct consequence of \cite{Rosen1965},
the existence of a Nash equilibrium is proved.
\end{IEEEproof}

Since $U_c(\cdot)$ is component-wise increasing, it follows that the maximum of (\ref{num-max1}) is obtained for some
$\gamma_c^\star$ such that $\overline{\Rm}(\gamma^\star_c,\gamma_{-c})$ is on the Pareto boundary of
$\Rc_c(\gamma_{-c})$. If the service rate function $R_k(g,\Hm, \chi, \Sigmam,\rv)$ is strictly increasing in
the power allocated to user $k$, then the Pareto boundary of $\Rc_c(\gamma_{-c})$ is achieved by
policies that satisfy $\trace\left ( \Sigmam_c[t] \right ) = \sum_{k\in \Sc_c[t]} P_{k,c}[t] = 1$ with probability 1.
In this case,  any Nash equilibrium $\{\gamma_c^\star : c = 1, \ldots, C\}$ must correspond to
scheduling policies that achieve the power constraint with equality for all BSs.

In Sections \ref{conventional} and \ref{incr-harq} we will focus on reference cell $c$, assuming that all other interfering cells apply a fixed
arbitrary policy $\gamma_{-c}$ (i.e., for fixed and known joint statistics of the ICI powers at all users of cell $c$).
We shall apply the theory developed in \cite{NeelyBook,Hooman-nonperfectCSI} and
provide a stochastic optimization algorithm that solves (\ref{num-max1}) to any desired level of
approximation, for any given ICI powers statistics.

\section{Scheduling with adaptive variable-rate coding and ARQ-LLC} \label{conventional}

From now on we shall assume Gaussian random coding and consider specific cases
of service rate functions. In this case, we define the $k$-th user mutual information function as
\begin{equation} \label{m-info}
I_k(g,\Hm, \chi, \Sigmam) = \log \left ( 1 +  \frac{g \left |\hv_{k}^\herm \vv_{k} \right |^2 P_{k}}{1 + \chi} \right )
\end{equation}
The mutual information at user $(k,c)$ receiver on slot $t$ is given by $I_{k,c}[t] \eqdef I_k(g_{k,c,c},\Hm_c[t], \chi_{k,c}[t], \Sigmam_c[t])$.
We approximate the decoding error probability by the corresponding
{\em information outage probability} (see \cite{Ozarow-Shamai-Wyner,biglieri1998fci} for the information-theoretic motivations underlying
this very common and very useful approximation). Namely, if the mutual information $I_{k,c}[t]$ is less than the scheduled
coding rate $r_{k,c}[t]$,  the decoder makes a decoding error with probability close to 1, while if $I_{k,c}[t] > r_{k,c}[t]$
the random coding average error probability is very close to 0. Therefore, for slot length $T$ large enough,
there exist ``good'' codes drawn from a Gaussian ensemble such that their block error probability
is close to the information outage probability $\PP(r_{k,c}[t] > I_{k,c}[t])$. In this case, the user $k$ service rate function
is given by ``outage rate'' function \cite{Hooman-nonperfectCSI}
\begin{equation} \label{outage-rate}
R_k(g,\Hm, \chi, \Sigmam,\rv) = r_k \times \mathbbm{1} \left \{ r \leq I_k(g,\Hm, \chi, \Sigmam) \right \}
\end{equation}
In order to obtain the desired near-optimal scheduling policy, we apply the framework of \cite{Hooman-nonperfectCSI}.
We define the {\em virtual queues}\footnote{It is important
to keep in mind that the virtual queues have nothing to do with the ARQ transmission buffers: they are used here as a tool to recursively update the weights of the
the scheduling policy.}  with buffer state $\Qm_c[t] = (Q_{1,c}[t], \ldots, Q_{K,c}[t])$ and {\em virtual arrival processes} $\Am_c[t] = (A_{1,c}[t], \ldots, A_{K,c}[t])$.
The virtual queues evolve according to the stochastic difference equations
\begin{equation} \label{queue-evo}
Q_{k,c}[t+1] = \max \left \{ 0 , Q_{k,c}[t] -  R_{k,c}[t] \right \} + A_{k,c}[t], \;\; k = 1,\ldots, K
\end{equation}
Then, we consider the {\em adaptive} policy defined by:
\begin{enumerate}
\item For any given $t$, let the transmit covariance matrix $\Sigmam_{c}[t]$ and the rate allocation vector $\rv_c[t]$  be the solution of
\begin{equation}  \label{stability-policy}
\begin{array}{ll}
\mbox{maximize} & \displaystyle{ \sum_{k = 1}^K \; Q_{k,c}[t]  \; \EE \left [  \left . r_{k,c}[t] \times \mathbbm{1}\left \{ r_{k,c}[t] \leq
I_k(g_{k,c,c},\Hm_c[t], \chi_{k,c}[t],\Sigmam_{k,c}[t]) \right \}  \right | \Hm_c[t] \right ]}   \\
\mbox{subject to} & \displaystyle{\trace\left ( \Sigmam_c[t] \right ) \leq 1, \;\;\;\; r_{k,c}[t]  \geq 0 \;\; \forall \; k}  \end{array}
\end{equation}
\item For suitable constants $V , A_{\max} > 0$, let the virtual arrival processes at time $t$ be given by  the solution of
\begin{equation} \label{flow-control}
\max_{0 \leq A_{k,c}[t] \leq A_{\max}, \; \forall k} \;\;\; V U_c(\Am_c[t]) - \sum_{k=1}^K A_{k,c}[t] Q_{k,c}[t]
\end{equation}
\item Update the virtual queues according to (\ref{queue-evo}),  with arrivals $\Am_c[t]$ given by (\ref{flow-control}) and service rates
$R_{k,c}[t]$ given by (\ref{outage-rate}) calculated for $\Sigmam_c[t]$ and $\rv_c[t]$
solutions of (\ref{stability-policy}).
\end{enumerate}
As stated in Theorem \ref{opt-lemma} below, the policy defined above achieves the optimal
point $\overline{\Rm}_c^\star$ solution of (\ref{num-max1}) within any desired accuracy, depending on
the constants $V$ and $A_{\max}$.  Neglecting the (small) degradation due to stochastic adaptation and
quantified by Theorem \ref{opt-lemma},  we shall refer to this policy as $\gamma^\star$.

\begin{thm} \label{opt-lemma}
Assume i.i.d. channels and fixed joint statistics of the ICI powers $\{\chi_{k,c} : k = 1,\ldots, K\}$.
Assume that $U_c(\cdot)$ is concave and entry-wise non-decreasing, and that there exists at least one point
$\rv \in \Rc_c(\gamma_{-c})$ with strictly positive entries such that  $U_c(\rv/2) > -\infty$.
Then, the scheduling policy $\gamma_c^\star$ defined above, for given constants $V>0$ and $A_{\max}>0$,
has the following properties:

(a) The utility achieved  by $\gamma^\star$ satisfies:
\begin{equation} \label{opt-a}
\liminf_{t \rightarrow \infty} U_c \left ( \frac{1}{t} \sum_{\tau=1}^{t} \Rm_c[\tau] \right )
\geq
U_c(\overline{\Rm}^\star(A_{\max})) - \kappa/V
\end{equation}
where
\begin{equation}  \label{C-constant}
\kappa \eqdef   \frac{1}{2}\left ( K A_{\max}^2 + \sum_{k=1}^K \EE\left [ \log^2\left (1 + \frac{g_{k,c,c} |\hv_{k,c,c}|^2}{1 + \chi_{k,c}} \right )\right ]  \right )
\end{equation}
and where $\overline{\Rm}_c^\star(A_{\max})$ denotes the solution of the problem (\ref{num-max1})
with the additional constraint $0 \leq \overline{R}_{k,c} \leq A_{\max}$ for all $k = 1,\dots,K$.

(b) For any point $\overline{\Rm}_c \in \Rc_c(\gamma_{c'} : c' \neq c)$ such that $0 \leq \overline{R}_{k,c} \leq A_{\max}$
for all $k$, and for any value $\beta \in [0,1]$ we have:
\begin{equation} \label{opt-b}
\limsup_{t \rightarrow\infty} \frac{1}{t} \sum_{\tau=1}^{t} \sum_{k=1}^K \overline{R}_{k,c}
\EE[ Q_{k,c}[\tau]] \leq \frac{\kappa + V[U_c(\overline{\Rm}_c^\star(A_{\max})) - U_c(\beta \overline{\Rm}_c)]}{1-\beta}
\end{equation}
Thus, all virtual queues $Q_{k,c}[t]$ are strongly stable.\footnote{
A discrete-time queue $Q_k[t]$ is \emph{strongly stable} if $\limsup_{t\rightarrow\infty}
\frac{1}{t} \sum_{\tau=1}^{t} \EE[Q_k[\tau]]  < \infty$. The system is strongly stable if all queues $k  = 1, \ldots, K$ are strongly stable.}
\end{thm}


\begin{IEEEproof}
The proof follows verbatim from the results in \cite{Hooman-nonperfectCSI} and it is not repeated here for brevity.
\end{IEEEproof}

As a corollary of Theorem \ref{opt-lemma}, if $A_{\max}$ is sufficiently large
such that  $A_{\max} \geq \overline{R}_{k,c}^\star$ for all $k$,
then $\gamma^\star_c$ satisfies
\begin{equation} \label{coro1}
\liminf_{t\rightarrow\infty} U_c \left ( \frac{1}{t} \sum_{\tau=1}^{t} \Rm_c[\tau]
\right )  \geq  U_c(\overline{\Rm}_c^\star) - \kappa/V.
\end{equation}
Hence, the control parameter $V$ can be chosen sufficiently large in order to make the achieved utility as close as desired to the optimal
value $U_c(\overline{\Rm}^\star)$ of problem (\ref{num-max1}). This comes with a tradeoff in the virtual queue average sizes that,
as seen from (\ref{opt-b}), grow linearly with $V$.  The virtual queue sizes represent the difference between the virtual
bits admitted into the queues and the actual bits transmitted, and thus affect the time-scales required for the time averages
to become close to their limiting values.

\subsection{Implementation} \label{impl-arq-llc}

The policy $\gamma^\star$ found before computes recursively the ``weights'' $\Qm_c[t]$ via (\ref{flow-control})
and (\ref{queue-evo}) and, for each $t$,  solves the weighted conditional average rate sum maximization (\ref{stability-policy}).
Problem (\ref{flow-control}) is a standard convex optimization problem the solution of which does not present any major conceptual difficulty and
is found in closed form for the important cases of proportional fairness and max-min fairness (see \cite{Hooman-nonperfectCSI}),
corresponding to the choices $U_c(\overline{\Rm}) = \sum_{k=1}^K \log \overline{R}_{k}$  and $U_c(\overline{\Rm}) = \min_k \overline{R}_k$,
respectively. In contrast, solving (\ref{stability-policy}) presents some difficulties.
Letting $F_{k,c}(\cdot)$ denote the marginal cdf of $\chi_{k,c}[t]$ and
using (\ref{m-info}), the objective function in (\ref{stability-policy}) can be rewritten as
\begin{eqnarray} \label{objective-stab}
& \sum_{k \in \Sc_c[t]} \; Q_{k,c}[t]   r_{k,c}[t]  \; \displaystyle{ F_{k,c} \left ( \frac{g_{k,c,c} \left |\hv^\herm_{k,c,c}[t] \vv_{k,c}[t] \right |^2 P_{k,c}[t]}{2^{r_{k,c}[t]} - 1}  - 1\right )} &
\end{eqnarray}
The optimization in (\ref{stability-policy}) is generally a non-convex problem that involves a
combinatorial search over all subsets $\Sc_c[t] \subseteq \{1, \ldots, K\}$ of cardinality $ \leq \min\{K,M\}$ and,
for each candidate subset,  the maximization of (\ref{objective-stab}) with respect to
$\rv_c[t]$ and the power allocation $\{P_{k,c}[t] : k \in \Sc_c[t]\}$.
Since this optimization may be difficult to compute, we propose the following suboptimal  low-complexity two-step approach:

Step 1) the active user subset and the corresponding power allocation are
selected by assuming deterministic ICI powers, equal to their mean value
$\overline{\chi}_{k,c}$. Under this assumption, the problem is reduced to the well-known
user selection with ZFBF,  that can be solved using standard techniques based on quasi-orthogonal
user selection and waterfilling  (e.g., \cite{DS05,Yoo-Goldsmith,Hoon-Caire-ISIT09}).

Step 2) for the transmit covariance $\Sigmam_c[t]$ obtained in step 1, (\ref{objective-stab}) is optimized
with respect to the rate allocation. This reduces to optimizing the outage rate separately
for each $k \in \Sc_{k,c}[t]$ by letting
\begin{equation} \label{rate-allocation-arq-llc}
r_{k,c}[t] = \arg \; \max_{r \geq 0 }  \displaystyle{ \left \{  r \; F_{k,c} \left ( \frac{g_{k,c,c} \left |\hv^\herm_{k}[t] \vv_{k,c,c}[t] \right |^2 P_{k,c}[t]}{2^{r} - 1}  - 1\right ) \right \}}
\end{equation}
where $g_{k,c,c} \left |\hv^\herm_{k,c,c}[t] \vv_{k,c}[t] \right |^2 P_{k,c}[t]$ is fixed by Step 1.

Notice that, both in the original problem and in the proposed low-complexity two-step approximated solution,
only the {\em marginal} statistics of the ICI powers $\{\chi_{k,c}[t] : k = 1,\ldots, K\}$ are relevant.
These marginal statistics can be measured by each user terminal individually and fed back to the BS scheduler by some very low-rate feedback scheme.

\section{Scheduling with incremental redundancy HARQ} \label{incr-harq}

If a genie provides the BS scheduler with the values of the the mutual information $\{I_{k,c}[t] : k = 1,\ldots, K\}$
in a non-causal fashion, just before the beginning of slot $t$, then the optimal rate allocation would be,
trivially, $r_{k,c}[t] = I_{k,c}[t]$ for all $k \in \Sc_c[t]$, yielding zero outage probability.
This ``genie-aided'' case was considered in \cite{Hooman-nonperfectCSI}  and referred to
as ``optimistic rate''  allocation, although no actual algorithm to approach the optimistic throughput was given.
Since for any non-negative random variable $I$ and $r > 0$ we have $\EE[r  \mathbbm{1}\{ r > I\}] \leq \EE[I]$, then
the optimistic service rates provide an upper bound to the throughput of any system with the same signaling scheme
(ZFBF and Gaussian codes) and given rate allocation.

In this section we show how to achieve the ``optimistic'' throughput without the aid of any genie.
As a preliminary step, let's consider the following {\em incremental redundancy} HARQ scheme.
The BS scheduler maintains a buffer of information packets for each user in the cell.
The size of user $(k,c)$ packets is equal to $b_{k,c}$ bits per packet.
Each packet is encoded into an infinite-length sequence of complex
symbols.\footnote{In practice, this rateless coding can be implemented  by using Raptor codes \cite{RaptorCode}.}
The encoded sequence is partitioned into blocks of length $T$ symbols.
At each slot $t$, the scheduling policy computes  $\Sigmam_c[t]$ according to
some rule to be found later. For all active users $k \in \Sc_c[t]$, if the most recent HARQ feedback message
from user $k$ is ``NACK'' (negative acknowledgement), then the first not-yet transmitted
coded block of the current  packet is transmitted. Otherwise, if  the most recent received HARQ feedback message
is ``ACK'' (positive acknowledgement),  then the current packet is removed from the transmission
buffer of user $k$ and the first coded block of next packet in the buffer is transmitted.
The $(k,c)$-th receiver stores in memory all the received slots for times $\{t : k \in \Sc_c[t]\}$
and attempts to decode the current packet at every newly received slot,
by using all the available received slots.
If decoding fails, NACK is sent back, otherwise ACK is sent back and the  decoder memory is reset.
Notice that the scheme does not require any genie-aided ``look-ahead'' of the instantaneous
ICI  power $\chi_{k,c}[t]$,  and makes use of time-invariant packet sizes $b_{k,c}$. These may differ from user to user
but are independent of $t$. For later use, we define the ``first-block coding rate'' as the
ratio $r_{k,c} = \frac{b_{k,c}}{T}$ bits/channel use.

Next, we describe a scheduling rule, denoted again by $\gamma_c^\star$, that
operates arbitrarily closely to the genie-aided throughput when
combined with the HARQ scheme described above.
At the end of each slot $t$, the active users $k \in \Sc_c[t]$ feed back both their
ACK/NACK message and the mutual information $I_{k,c}[t]$ ``seen'' at their receiver.
Then, $\gamma_c^\star$ coincides with what given in Section \ref{conventional}, after the following two changes.
1) The virtual queues evolution equation (\ref{queue-evo}) is replaced by
\begin{equation} \label{queue-evo1}
Q_{k,c}[t+1] = \max \left \{ 0 , Q_{k,c}[t] -  I_{k,c}[t] \right \} + A_{k,c}[t], \;\;\; \forall \; k
\end{equation}
2) The transmitter optimization (\ref{stability-policy}) is replaced by
\begin{equation}  \label{stability-policy1}
\begin{array}{ll}
\mbox{maximize} &  \displaystyle{\sum_{k =1}^K \; Q_{k,c}[t]  \; \EE \left [ \left .  I_k(g_{k,c,c}, \Hm_c[t],\chi_{k,c}[t], \Sigmam_c[t]) \right | \Hm_c[t] \right ]}   \\
\mbox{subject to} & \displaystyle{\trace\left ( \Sigmam_c[t] \right ) \leq 1} \end{array}
\end{equation}
In brief, the scheduler updates recursively its weights $\Qm_c[t]$ and computes the
transmitted signal covariance $\Sigmam_c[t]$  according to (\ref{stability-policy1}),
as if it was operating on a virtual ``genie-aided'' system with instantaneous
service rates $I_{k,c}[t]$.
The throughput region of the virtual genie-aided system,
denoted by $\Rc_c^{\rm genie}(\gamma_{-c})$,  is given by (\ref{ergodic-region}), after replacing the general rate function $R_k(\cdots)$
with the mutual information function $I_k(\cdots)$ defined in (\ref{m-info}).
The performance of $\gamma_c^\star$ for the genie-aided system is again given
by Theorem \ref{opt-lemma}, where $\Rm_c[\tau]$ in (\ref{opt-a}) is replaced by the vector of mutual informations
$\Id_c[\tau] = (I_{1,c}[\tau, \ldots, I_{K,c}[\tau])$ and where  $\overline{\Rm}^\star(A_{\max})$ denotes the solution of (\ref{num-max1}) when
$\Rc_c(\gamma_{-c})$ is replaced by $\Rc_c^{\rm genie}(\gamma_{-c})$, with the additional constraint
$0 \leq \overline{R}_{k,c} \leq A_{\max}$ for all $k = 1,\dots,K$.

For sufficiently large $A_{\max}$, $\gamma^\star_c$ yields:
\begin{equation} \label{coro2}
\liminf_{t\rightarrow\infty} U_c \left ( \frac{1}{t} \sum_{\tau=1}^{t} \Id_c[\tau] \right )  \geq  U_c(\overline{\Rm}_c^{{\rm genie},\star}) - \kappa/V,
\end{equation}
where $\overline{\Rm}_c^{{\rm genie},\star}$ is the utility-maximizing throughput point in the region $\Rc_c^{\rm genie}(\gamma_{-c})$.
At this point, it remains to be shown that the combination of the policy $\gamma_c^\star$
with the incremental redundancy HARQ scheme yields a network utility as close as desired to the limit in (\ref{coro2}). This is shown by the following:

\begin{thm} \label{harq}
Let $\overline{\Rm}^{{\rm harq}, \star}_c = (\overline{R}_{1,c}^{{\rm harq},\star}, \ldots, \overline{R}_{K,c}^{{\rm harq},\star})$
denote the throughput achievable by the incremental redundancy HARQ protocol under scheduling policy $\gamma_c^\star$
defined above.  For each user $(k,c)$ and $\epsilon_{k,c} > 0$  there exists a sufficiently large first-block rate $r_{k,c}$
such that $\overline{R}_{k,c}^{{\rm harq},\star} \geq (1 -  \epsilon_{k,c}) \overline{R}_{k,c}^{{\rm genie},\star}$.
\end{thm}

\begin{figure}
   \centering
   \includegraphics[width=14cm]{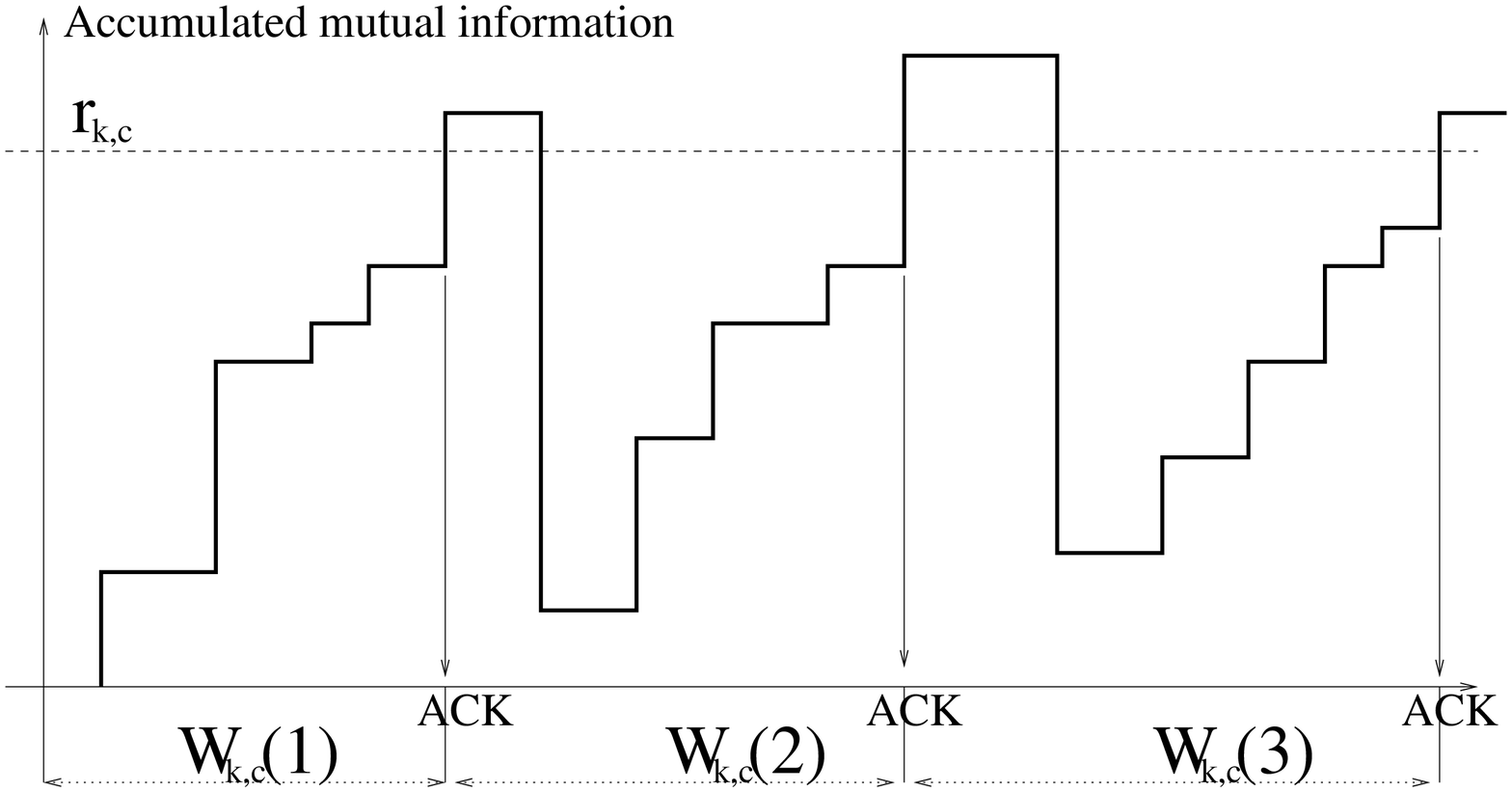}
   \caption{Qualitative plot of the mutual information level-crossing process that determines the decoding events of the HARQ protocol.
   The jumps of the accumulated mutual information process correspond to slot times at which user $(k,c)$ is active.}
  \label{m-info-level-crossing}
\end{figure}

\begin{IEEEproof}
Consider user $(k,c)$.  Following the argument in \cite{caire-tuninetti}, we can model the event of successful decoding
as a ``mutual information level-crossing event''.  Suppose that the transmission of the current packet
for user $(k,c)$ starts at slot $t_{\rm start}$ (i.e., an ACK was fed back at slot time $t_{\rm start} - 1$).
Then, the current packet can be successfully decoded
at slot $t \geq t_{\rm start}$ if $\sum_{\tau=t_{\rm start}}^t I_{k,c}[\tau] \geq r_{k,c}$.
Otherwise, a decoding error occurs with very high probability.  As shown in \cite{caire-tuninetti,Forney-1968},
the probability of undetected decoding error vanishes exponentially with $T$. Therefore, in the regime of large $T$,
if  $\sum_{\tau=t_{\rm start}}^t I_{k,c}[\tau] < r_{k,c}$ the decoding error is detected with arbitrarily high probability
and a NACK is sent back. Fig. \ref{m-info-level-crossing} shows, qualitatively, the mutual information level-crossing and the
corresponding successful decoding  events of the $(k,c)$ decoder. Notice that the mutual information increment is non-negative,
and it is exactly zero for all $t$ such that $k \notin \Sc_c[t]$, i.e., when user $(k,c)$ is not scheduled.

Consider the transmission of a long sequence of packets. Without loss of generality, assume that the system starts at time $t_{\rm start} = 1$,
denote by $N_{k,c}[t]$ the number  of successful decoding events  of decoder $(k,c)$ up to time $t$ and
let $W_{k,c}(1), W_{k,c}(2), \ldots, W_{k,c}(N_{k,c}[t])$ denote the corresponding ``inter-ACK'' times (see Fig. \ref{m-info-level-crossing}).
Since at each successful decoding a ``reward'' of $r_{k,c}$ bit per channel use is delivered
to the destination,  the throughput of the HARQ protocols is given by
\begin{equation} \label{harq-throughput1}
\overline{R}_{k,c}^{{\rm harq},\star} = \lim_{t \rightarrow \infty}
\frac{r_{k,c} N_{k,c}[t]}{\sum_{n=1}^{N_{k,c}[t]} W_{k,c}(n) + \Delta_{k,c}[t]}
\end{equation}
where $\Delta_{k,c}[t] = t - \sum_{n=1}^{N_{k,c}[t]} W_{k,c}(n)$ denotes the difference between the
current time $t$ and the time at which the $N_{k,c}[t]$-th successful decoding occurred.
Under the assumptions of this paper,  the system with
HARQ protocol and scheduling policy $\gamma_c^\star$ evolves
according to a discrete-time, continuous-valued vector Markov process with state
given by  $\Qm_c[t]$ and by the vector of accumulated mutual  informations at each receiver.
Since the virtual queues are strongly stable (see Theorem \ref{opt-lemma})
and the accumulated mutual informations are bounded in  $[0,r_{k,c}]$,
the process is stationary and ergodic.
Therefore, the limit in (\ref{harq-throughput1}) holds almost surely, and can be explicitly
computed as follows:
\begin{eqnarray} \label{harq-throughput2}
\overline{R}_{k,c}^{{\rm harq},\star}
& = & \lim_{t \rightarrow \infty}
\frac{r_{k,c} }{\frac{1}{N_{k,c}[t]} \sum_{n=1}^{N_{k,c}[t]} W_{k,c}(n) + \frac{\Delta_{k,c}[t]}{N_{k,c}[t]}} \nonumber \\
& = &
\frac{r_{k,c} }{\lim_{t \rightarrow \infty}  \frac{1}{N_{k,c}[t]} \sum_{n=1}^{N_{k,c}[t]} W_{k,c}(n) + \lim_{t \rightarrow \infty}  \frac{\Delta_{k,c}[t]}{N_{k,c}[t]}} \nonumber \\
& = &
\frac{r_{k,c} }{\EE[W_{k,c}]}
\end{eqnarray}
where $W_{k,c}$ is an integer-valued random variable with the same marginal distribution of the inter-ACK times.

In order to determine $\EE[W_{k,c}]$, consider the case $t_{\rm start} = 1$ and define the event
\begin{equation} \label{level-crossing}
\Ac_{k,c}[t] = \left \{ \sum_{\tau=1}^t I_k[\tau] \; \leq \; r_{k,c} \right \}
\end{equation}
Since the accumulated mutual information between two ACKs is non-decreasing, the following nesting condition holds:
\[ \Ac_{k,c}[t] \subseteq \Ac_{k,c}[t-1], \;\; \forall \; t \]
where $\Ac_{k,c}[0] = \{0 \leq r_{r,c}\}$ has probability 1.  It follows that
\[ \PP(W_{k,c} = t) = \PP(\Ac_{k,c}[t-1], \overline{\Ac_{k,c}[t]}) = \PP(\Ac_{k,c}[t-1]) - \PP(\Ac_{k,c}[t]), \]
yielding the average inter-ACK time in the form
\begin{eqnarray} \label{renewal}
\EE[W_{k,c}] & = & \sum_{t=1}^\infty t \PP(W_{k,c} = t) \nonumber \\
& = & 1 + \sum_{t=1}^\infty \PP(\Ac_{k,c}[t])
\end{eqnarray}
Owing to the complete formal analogy of results (\ref{harq-throughput2}) and (\ref{renewal}) with
the throughput of HARQ considered in \cite{caire-tuninetti}), we can directly apply the limit proved
in \cite{caire-tuninetti}:
\footnote{This result is indeed quite intuitive: when $r_{k,c}$ becomes large, then $\EE[W_{k,c}]$ increases. Therefore, the
accumulated mutual information divided by the number of slots $\frac{1}{W_{k,c}} \sum_{\tau=1}^{W_{k,c}}
I_k \left ( g_{k,c,c}, \Hm_c[\tau], \chi_{k,c}[\tau], \Sigmam_c[\tau] \right )$ converges to an ensemble average.
It follows that in this limit the level crossing condition tends to become deterministic, and satisfies (approximately)
\[  \sum_{\tau=1}^{W_{k,c}}   I_k \left ( g_{k,c,c}, \Hm_c[\tau], \chi_{k,c}[\tau], \Sigmam_c[\tau] \right ) = W_{k,c} r_{k,c} \]
Of course, this argument can be made rigorous by following in the footsteps of \cite{caire-tuninetti}.}
\begin{equation}  \label{renewal-limit}
\lim_{r_{k,c} \rightarrow \infty} \frac{r_{k,c}}{\EE[W_{k,c}]} = \EE \left [  I_k \left ( g_{k,c,c}, \Hm_c, \chi_{k,c}, \Sigmam_c \right ) \right ]
\end{equation}
In particular, as $r_{k,c} \rightarrow \infty$ the average inter-ACK time $\EE[W_{k,c}]$ diverges to infinity linearly with $r_{k,c}$.
The analysis in \cite{caire-tuninetti} shows that, for any $\eta_{k,c} > 0$,
\begin{equation} \label{tuni}
\overline{R}^{{\rm harq},\star}_{k,c} \geq (1 - \eta_{k,c}) \EE \left [  I_k \left ( g_{k,c,c}, \Hm_c, \chi_{k,c}, \Sigmam_c \right ) \right ]
\end{equation}
for all sufficiently large $r_{k,c}$.

The proof of Lemma \ref{harq} is finally concluded by  combining  the result (\ref{renewal-limit}) with (\ref{coro2}).
By stationarity and ergodicity, under $\gamma_c^\star$ we have that
\[ \lim_{t \rightarrow \infty} \frac{1}{t} \sum_{\tau=1}^t I_{k,c}[\tau] = \EE \left [  I_k \left ( g_{k,c,c},  \Hm_c, \chi_{k,c}, \Sigmam_c \right ) \right ] \]
holds almost surely. Since $U_c(\cdot)$ is component-wise increasing, (\ref{coro2}) implies that for any $\delta_{k,c} > 0$ there exist
sufficiently large $A_{\max}$ and $V$ for which
\begin{equation} \label{monotonic}
\EE \left [  I_k \left ( g_{k,c,c}, \Hm_c, \chi_{k,c}, \Sigmam_c \right ) \right ] \geq (1 - \delta_{k,c}) \overline{R}_{k,c}^{{\rm genie},\star}
\end{equation}
By letting $(1 - \epsilon_{k,c}) = (1 - \eta_{k,c}) (1 - \delta_{k,c})$ and using (\ref{tuni}) and (\ref{monotonic}) Theorem \ref{harq} is proved.
\end{IEEEproof}

From the above proof it follows that the delay-throughput
operating point of the incremental redundancy HARQ protocol can be chosen individually for each user by setting
the threshold value $r_{k,c}$ (or, equivalently, the size $b_{k,c}$  of the information packets).
By making $r_{k,c}$ large, the average decoding delay $D_{k,c} = \EE[W_{k,c}]$ becomes large and the throughput
approaches $\overline{R}_{k,c}^{{\rm genie},\star}$.

Also, we wish to stress the difference between the ARQ-LLC scheme described in Section \ref{model} and the incremental-redundancy
HARQ protocol illustrated in this section. The ARQ-LLC protocol makes use of adaptive variable-rate coding at the physical layer,
and removes or keeps in the transmission buffer packets of information bits of variable size
$b_{k,c}[t] = T r_{k,c}[t]$.  In contrast, the HARQ protocol  make use of a fixed packet size $b_{k,c}$ (equivalent to fixed first-block
rate $r_{k,c}$), but the effective service rate is adaptive by varying the decoding delay through the ACK/NACK mechanism.

\subsection{Implementation} \label{impl-harq}

The scheme previously proposed requires that each active user, at the end of each slot $t$, feeds back a message formed by
one bit for ACK/NACK and by the value of $I_{k,c}[t]$ or, equivalently, the value of $\SINR_{k,c}[t]$.
We notice that feeding back the instantaneous SINR is widely proposed in the literature on opportunistic
downlink scheduling \cite{hassibi-sharif,yoo2007jsac} and it is referred to as {\em Channel Quality Indicator} (CQI).
However, in the current literature the CQI is relative to the {\em current} slot, and it is used to select users and
allocate the rate of a variable-rate coding scheme. In contrast, here the CQI refers to the {\em past} slot, and it is used to update the
scheduler weights  according to (\ref{queue-evo1}).

Denoting again by $F_{k,c}(\cdot)$ the marginal cdf of $\chi_{k,c}[t]$,  the objective function in (\ref{stability-policy1}) can be rewritten as
\begin{eqnarray} \label{objective-stab1}
& \sum_{k \in \Sc_c[t]} \; Q_{k,c}[t]   \displaystyle{ \int_0^\infty \log \left (1 + \frac{g_{k,c,c} \left |\hv_{k,c,c}^\herm[t] \vv_{k,c}[t] \right |^2 P_{k,c}[t]}{1 + z} \right )} \; dF_{k,c}(z)
\end{eqnarray}
While for any fixed user subset $\Sc_c[t]$ the maximization of (\ref{objective-stab1}) with respect to the powers $\{P_{k,c}[t] : k \in \Sc_c[t]\}$ is a convex
problem, the solution is not generally given by the simple waterfilling formula and it may be difficult to compute since
the cdfs $F_{k,c}(\cdot)$ are typically not known in closed form.
A near-optimum  low-complexity approximation consists  of choosing $\Sigmam_c[t]$ that maximizes the objective function lower bound
\begin{eqnarray} \label{objective-stab2}
& \sum_{k \in \Sc_c[t]} \; Q_{k,c}[t]   \; \displaystyle{\log \left (1 + \frac{g_{k,c,c} \left |\hv_{k,c,c}^\herm[t] \vv_{k,c}[t] \right |^2 P_{k,c}[t]}{1 + \overline{\chi}_{k,c}} \right )}
\end{eqnarray}
obtained by applying Jensen's inequality to (\ref{objective-stab1}).
Notice that the maximization of (\ref{objective-stab2}) with respect to the transmit covariance matrix
coincides with step 1 in the low-complexity approximation of the variable-rate coding/ARQ-LLC case of
Section \ref{impl-arq-llc},  and can be solved efficiently using the methods
in \cite{DS05,Yoo-Goldsmith,Hoon-Caire-ISIT09}.

\subsection{Extremal ICI distributions} \label{extremal}


The throughput performance of the HARQ scheme depends on the statistics of the ICI powers, which in turns
depend on the scheduling policies $\gamma_{-c}$ at the interfering BSs.
In this section we find {\em extremal} marginal statistics for the ICI powers that provide non-trivial inner
and outer bounds to $\Rc_c^{\rm genie}(\gamma_{-c})$ that are independent of $\gamma_{-c}$.
Here we drop the slot index $t$ since all processes are stationary. We start with the following:

\begin{lemma} \label{extremal-lemma}
For all feasible policies $\gamma_{c'} : c' \neq c$ that satisfy the input power constraint with equality and
for all users $k = 1,\ldots, K$, we have
\begin{eqnarray} \label{jensen}
& \displaystyle{\EE[ I_k(g_{k,c,c},\Hm_c,\overline{\chi}_{k,c},\Sigmam_c) ] \leq
\EE[ I_k(g_{k,c,c},\Hm_c,\chi_{k,c},\Sigmam_c) ] \leq
\EE[ I_k(g_{k,c,c},\Hm_c,\widetilde{\chi}_{k,c},\Sigmam_c) ]} &
\end{eqnarray}
where $\overline{\chi}_{k,c} = \EE[\chi_{k,c}] = \sum_{c'\neq c} g_{k,c,c'}$ and where
$\widetilde{\chi}_{k,c} = \sum_{c' \neq c} g_{k,c,c'} \left |\hv^\herm_{k,c,c'} \vv_{1,c'} \right |^2$
is the ICI power at the $(k,c)$ receiver when all interfering BSs $c' \neq c'$ schedule a single user in their own cell.
\end{lemma}

\begin{IEEEproof}
The first inequality (lower bound) follows immediately from Jensen's inequality applied to the convex function $f(x) = \log(1 + \frac{a}{b + x})$ with $a,b > 0$, and by the fact that, by assumption, the interfering BSs use all their available power.
In order to show the second inequality (upper bound), we use (\ref{tx-cov}) in (\ref{inst-ici}) and write
$\chi_{k,c} = \sum_{c'\neq c} g_{k,c,c'} \sum_{j \in \Sc_{c'}} \alpha_{k,c,c',j} P_{j,c'}$,
where $\alpha_{k,c,c',j} \eqdef |\hv_{k,c,c'}^\herm \vv_{j,c'}|^2$ are random variables independent of
the SINR numerator $|\hv_{k,c,c}^\herm \vv_{k,c}|^2 P_{k,c}$.
Since the ZFBF steering vectors $\vv_{j,c'}$ have unit norm and are independent of
$\hv_{k,c,c'}$, the variables $\alpha_{k,c,c',j}$ are marginally identically distributed
as central chi-squared with 2 degrees of freedom \cite{Probability-RandomProcesses}.
Also, notice that the $\alpha_{k,c,c',j}$'s are statistically dependent for the same
index $c'$, while  $\{\alpha_{k,c,c',j} : j \in \Sc_{c'}\}$ and $\{\alpha_{k,c,c'',j} : j \in \Sc_{c''}\}$ are group-wise mutually independent
for $c' \neq c''$.  By assumption,  $\sum_{j \in\Sc_{c'}} P_{j,c'} = 1$ for all $c'$.
Therefore,  $\sum_{j \in \Sc_{c'}} \alpha_{k,c,c',j} P_{j,c'}$ is a convex combination of identically distributed, possibly dependent, random variables.
The second inequality in (\ref{jensen}) follows by repeated application of Jensen's inequality. Choose $c'' \neq c$. Then, using (\ref{m-info}), we have
\begin{eqnarray} \label{jensen1}
& \displaystyle{\EE \left [ \log \left ( 1 + \frac{g_{k,c,c} \left |\hv_{k,c,c}^\herm \vv_{k,c} \right |^2 P_{k,c}}{1 + \chi_{k,c}} \right ) \right ]}
\leq &  \nonumber \\
& \displaystyle{ \sum_{j \in \Sc_{c''}} P_{j,c''}
\EE \left [ \log \left ( 1 + \frac{g_{k,c,c} \left |\hv_{k,c,c}^\herm \vv_{k,c} \right |^2 P_{k,c}}{1 + g_{k,c,c''} \alpha_{k,c,c'',j} +
\sum_{c' \neq c, c''} g_{k,c,c'} \sum_{j \in \Sc_{c'}} \alpha_{k,c,c',j} P_{j,c'}} \right ) \right ]} = & \nonumber \\
& \displaystyle{ \EE \left [ \log \left ( 1 + \frac{g_{k,c,c} \left |\hv_{k,c,c}^\herm \vv_{k,c} \right |^2 P_{k,c}}{1 +
g_{k,c,c''} \alpha_{k,c,c'',1} +  \sum_{c' \neq c, c''} g_{k,c,c'} \sum_{j \in \Sc_{c'}} \alpha_{k,c,c',j} P_{j,c'}} \right ) \right ]} &
\end{eqnarray}
where the equality in (\ref{jensen1}) follows from the fact that the $\alpha_{k,c,c'',j}$'s are identically distributed with respect to the index $j$.
Next, pick $c''' \neq c, c''$, and apply the same steps to the last line of (\ref{jensen1}).
After eliminating all convex combinations, the final upper bound coincides with the right most term in (\ref{jensen}).
\end{IEEEproof}
As a corollary, we have the following interesting ``robustness'' result:

\begin{thm} \label{robustness}
For any choice of the scheduling policies $\gamma_{-c}$ that satisfy the input power constraint with equality, we have
\begin{equation} \label{bounds}
\overline{\Rc}_c \subseteq \Rc_c^{\rm genie}(\gamma_{-c}) \subseteq \widetilde{\Rc}_c^{\rm genie}
\end{equation}
where  $\overline{\Rc}_c$ is the region with deterministic ICI powers $\{\overline{\chi}_{k,c}\}$,\footnote{Notice that if the
ICI powers were deterministic, then no genie or HARQ is needed and the system reduces to a collection of isolated cells,
where each cell $c$ has modified channel path gain coefficients $\overline{g}_{k,c,c} = \frac{g_{k,c,c}}{1 + \overline{\chi}_{k,c}}$.
In this case, the throughput region $\overline{\Rc}_c$ is achieved by the standard scheduling/resource allocation schemes with perfect state information and
zero outage probability.}
and  where  $\widetilde{\Rc}_c^{\rm genie}$ is the region corresponding to random ICI powers $\{\widetilde{\chi}_{k,c}\}$. Furthermore,
the gap between the inner and outer bounds in (\ref{bounds}) is bounded by a constant that does not depend on the channel path coefficients.
\end{thm}

\begin{IEEEproof}
The proof (\ref{bounds}) follows directly  as a consequence of Lemma \ref{extremal-lemma}.
In order to show the bounded gap, we have to find some constant $\Delta$,  independent of $\{g_{k,c,c'}\}$,
such that  $\max\{ \rv - \Delta \onev, \zerov\} \in  \overline{\Rc}_c$  for all points $\rv \in \widetilde{\Rc}_c^{\rm genie}$.
To this purpose, pick a point $\rv \in \widetilde{\Rc}_c^{\rm genie}$ corresponding to some feasible scheduling policy
$\gamma_c$ for the genie-aided system.
Applying the {\em same} sequence of input covariance matrices as determined by $\gamma_c$, to the
system with deterministic ICI powers, we certainly find a point $\overline{\Rm}_c(\gamma_c) \in \overline{\Rc}_c$.
Consider the throughput of the $k$-th user and let for convenience $A \eqdef g_{k,c,c} \left |\hv_{k,c,c}^\herm \vv_{k,c} \right |^2 P_{k,c}$.
Then, by applying Jensen's inequality we have
\begin{eqnarray} \label{gap}
& \EE \left [ \left . \log \left ( 1 + \frac{A}{1 +  \sum_{c' \neq c} g_{k,c,c'}  \alpha_{k,c,c',1} } \right ) \right | A \right ] - \log \left ( 1 + \frac{A}{1 +  \sum_{c' \neq c} g_{k,c,c'}  } \right ) & \nonumber \\
& \leq & \nonumber \\
& \log \left ( 1 + \sum_{c'\neq c} g_{k,c,c'} \right ) - \EE \left [ \log \left ( 1 + \sum_{c'\neq c} g_{k,c,c'} \alpha_{k,c,c',1} \right ) \right ] &
\end{eqnarray}
The RHS in the above inequality is easily seen to be non-negative and component-wise increasing with respect to
any coefficient $g_{k,c,c'}$.
Therefore, its maximum is obtained in the limit for all $g_{k,c,c'} \rightarrow \infty$ (in passing, we notice that this corresponds to considering
the interference-limited regime where SNR $\rightarrow \infty$). In order to see that this limit is finite, let $g_{\max} = \max g_{k,c,c'}$,
then we have
\begin{eqnarray}
\mbox{RHS of (\ref{gap})}
& \leq & \log \left ( 1 + (C-1) g_{\max} \right ) - \EE \left [ \log \left ( 1 +  g_{\max} \sum_{c'\neq c} \alpha_{k,c,c',1} \right ) \right ] \nonumber \\
& \leq & - \EE \left [ \log \left ( \frac{1}{C-1} \sum_{c'\neq c} \alpha_{k,c,c',1} \right ) \right ] \label{ziogap} \\
& \leq & - \EE \left [ \log \left ( \alpha_{k,c,c',1} \right ) \right ] \label{ziogap1} \\
& \leq & \gamma/\ln(2) \label{ziogap2}
\end{eqnarray}
where (\ref{ziogap}) follows by  letting $g_{\max} \rightarrow \infty$, (\ref{ziogap1}) follows by applying Jensen's inequality to the convex function $-\log x$ and
(\ref{ziogap2}) follows by using the fact that $\alpha_{k,c,c',1}$ is chi-squared with 2 degrees of freedom, and
using the limit $\lim_{\epsilon \downarrow 0} \int_{\epsilon}^\infty \ln x e^{-x} dx = -\gamma$, where $\gamma$ denotes
the Euler-Mascheroni constant \cite{abramowitz1965handbook}.
\end{IEEEproof}

Theorem \ref{robustness} has the following interesting consequence: consider the
multi-cell decentralized scheduling game under the proposed incremental redundancy HARQ scheme,
achieving the genie-aided throughput region in each cell. The performance of any given cell $c$ (in terms of its
network utility value) at any Nash equilibrium $(\gamma_1^\star, \ldots, \gamma_C^\star)$
is bounded below and above by the solutions of (\ref{num-max1}) when $\Rc_c^{\rm genie}(\gamma_{-c}^\star)$ is replaced by
$\overline{\Rc}_c$ and $\widetilde{\Rc}_c^{\rm genie}$, respectively. This follows from the fact that, as argued at
the end of Section \ref{model}, all Nash equilibria must achieve the power constraints with equality at each BS.\footnote{Notice that the mutual information function is strictly increasing with the SINR.}

\section{Numerical results} \label{results}

We considered a simple one-dimensional cellular layout with unit width cells
arranged on a line. BSs are located at integer positions $c \in \ZZ$.
In each cell $c$, users are placed on a uniform grid in positions
$u(k, c) = (2k - K - 1)/(2K) + c$, for $k = 1, \cdots , K$.
The channel path gains are given by $g_{k,c,c'} = \frac{G_0}{1 + (|u(k, c)-c'|_C/\delta)^\nu}$,
where the modulo-$C$ distance $|u - c|_C = \min\{|u - c + zC| : z \in \ZZ\}$ induces a torus topology that eliminates
border effects and where $\nu$ and $\delta$ are the propagation exponent and the 3dB breakpoint distance,
respectively, and $G_0$ determines the received SNR at the cell edge \cite{goldsmith2005wc}.
We present results for a system with $C = 18$ cells,
$M = 2$ antennas  per BS, $K = 36$ users per cell and parameters $G_0 = 60$dB, $\alpha=3.0$ and $\delta=0.05$.
For the implementation of the policy $\gamma_c^\star$ we chose parameters
$A_{\max} = 50$, $V=50$ and suboptimal low-complexity approximations
as explained in Sections \ref{impl-arq-llc} and \ref{impl-harq}, respectively.
As for the network utility functions, we considered both  proportional fairness and max-min fairness
(see Section \ref{impl-arq-llc} and \cite{Mo-Walrand,Hooman-nonperfectCSI} and references therein).
In order to gather the ICI statistics, we run the same scheduling algorithm in all BSs and measure the empirical cdf of the ICI power
at each user location in the reference cell $c = 0$ (since the system is completely symmetric, all cells see the same ICI statistics).

\begin{figure}
   \centering
   \includegraphics[width=14cm]{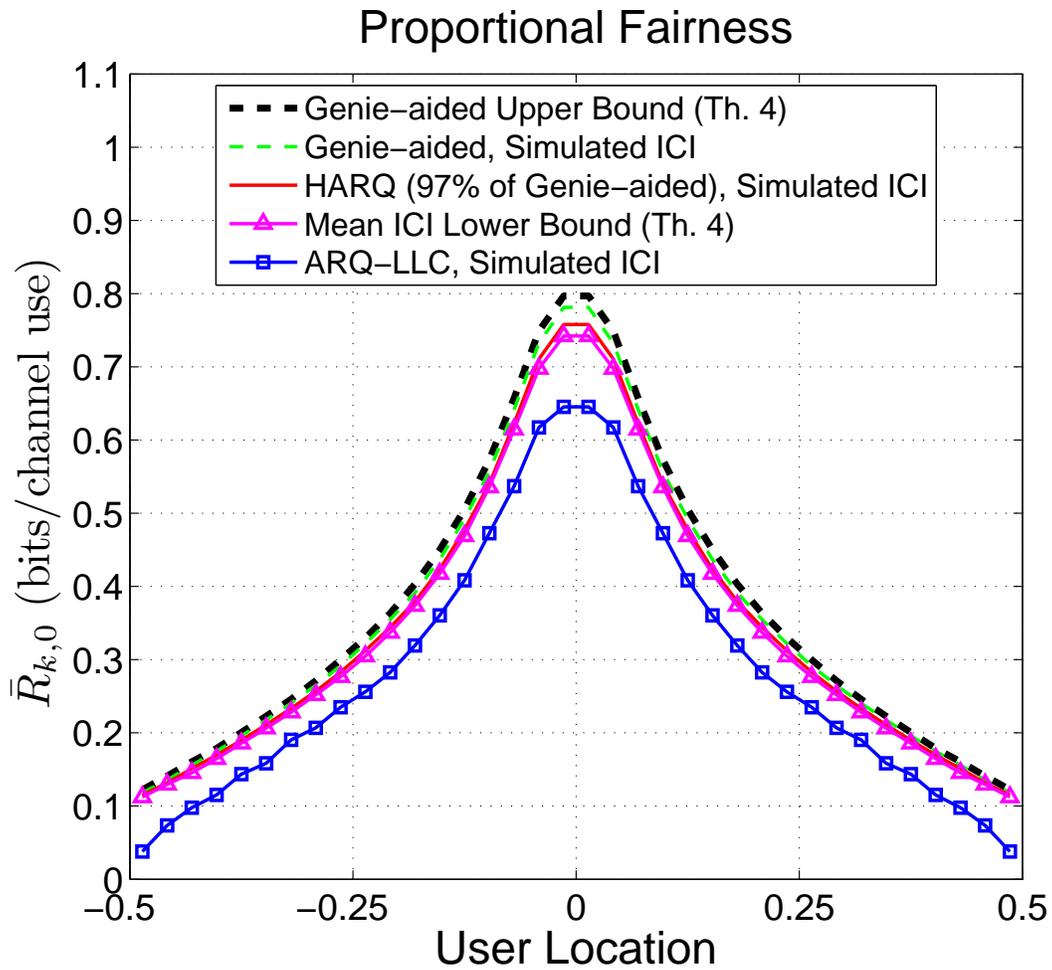}
   \caption{Average-throughput, proportional fairness.}
  \label{fig:AvgRate-UL-PFS}
 \end{figure}

\begin{figure}
   \centering
   \includegraphics[width=14cm]{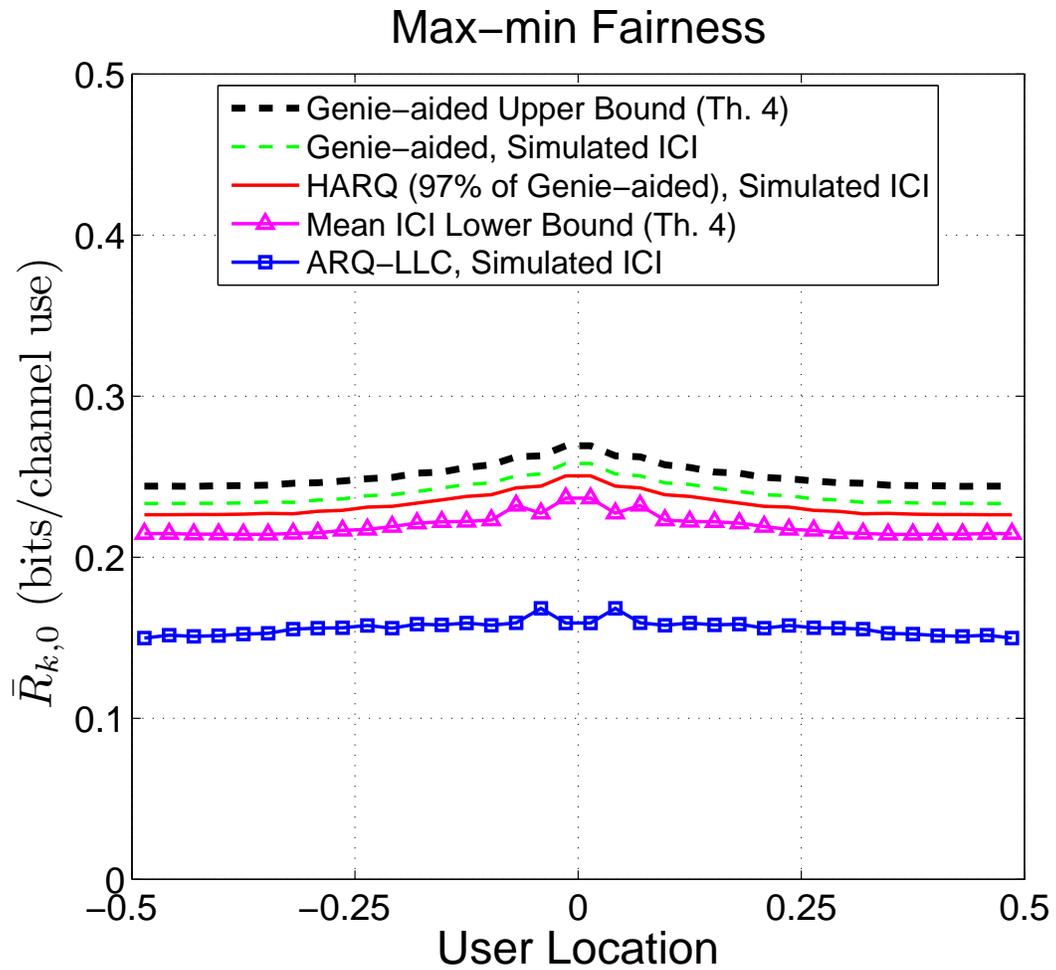}
   \caption{Average-throughput, max-min fairness.}
  \label{fig:AvgRate-UL-HFS}
 \end{figure}

Figs.~\ref{fig:AvgRate-UL-PFS} and~\ref{fig:AvgRate-UL-HFS} compare user throughputs in cell
$c=0$ under proportional fairness and max-min fairness, respectively.
Thick dashed lines illustrate the throughput upper bounds of Theorem \ref{robustness}.
Thin dashed lines  correspond to the actual ``genie-aided'' rates achievable by the proposed HARQ scheme
in the limit of infinite decoding delay. Solid lines show the throughput achieved by the HARQ scheme operating at
finite average decoding delay for all users,  by setting the parameters $\{r_{k,0}\}$ such that each user achieves 97\% of the genie-aided
rates (infinite delay). The ``triangle'' marks indicate the throughput lower bounds of Theorem \ref{robustness}.
Finally, the ``square'' marks indicate the throughputs achieved by the conventional adaptive variable-rate coding with ARQ-LLC.
We observe that  under both fairness objective functions, the throughputs achieved by HARQ
achieve a gain of more than 100\% for the users at the edge of the cell in the proportional fairness case, and a
throughput gain of more than 40\% for all users in the max-min fairness case, with respect to the ARQ-LLC scheme.

\begin{figure}
  \centering
  \includegraphics[width=14cm]{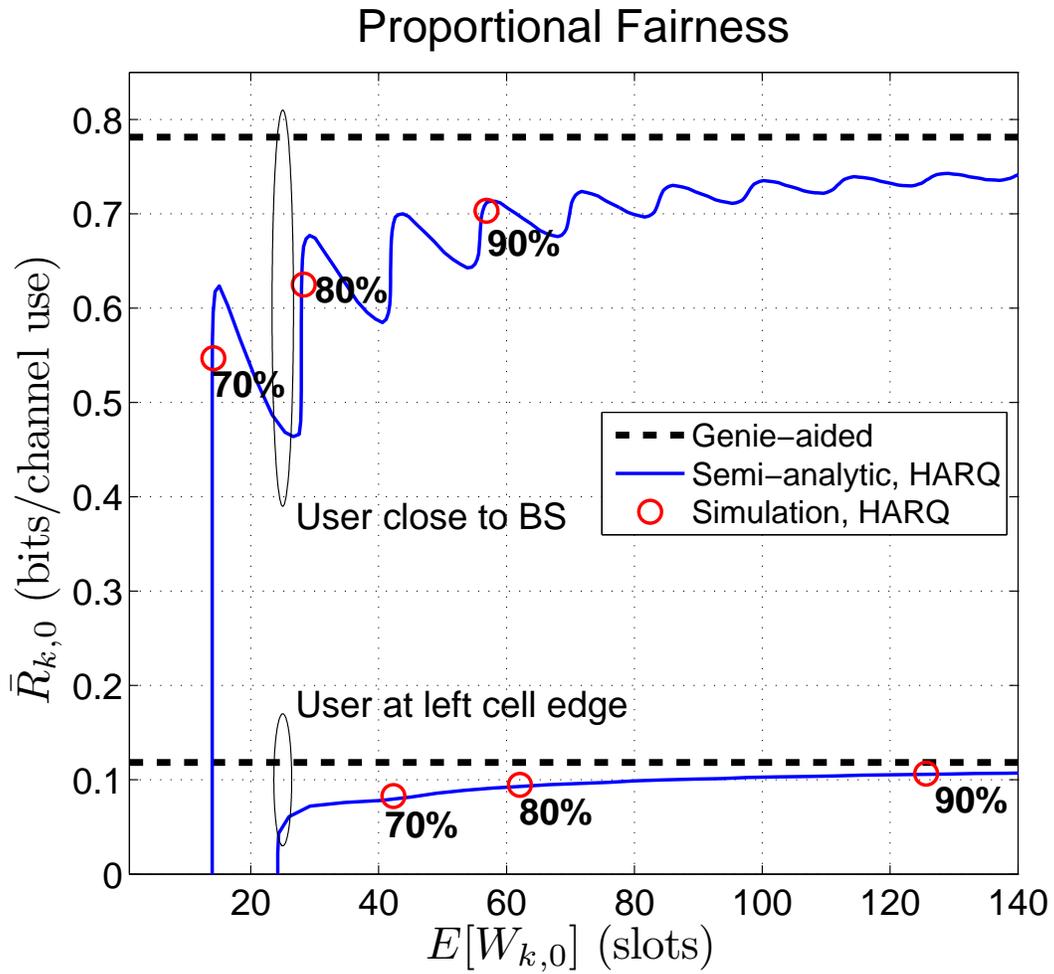}
  \caption{Average rate vs. decoding delay with proportional fairness for two sample users.}
 \label{fig:rate_delay_PFS}
\end{figure}

\begin{figure}
  \centering
  \includegraphics[width=14cm]{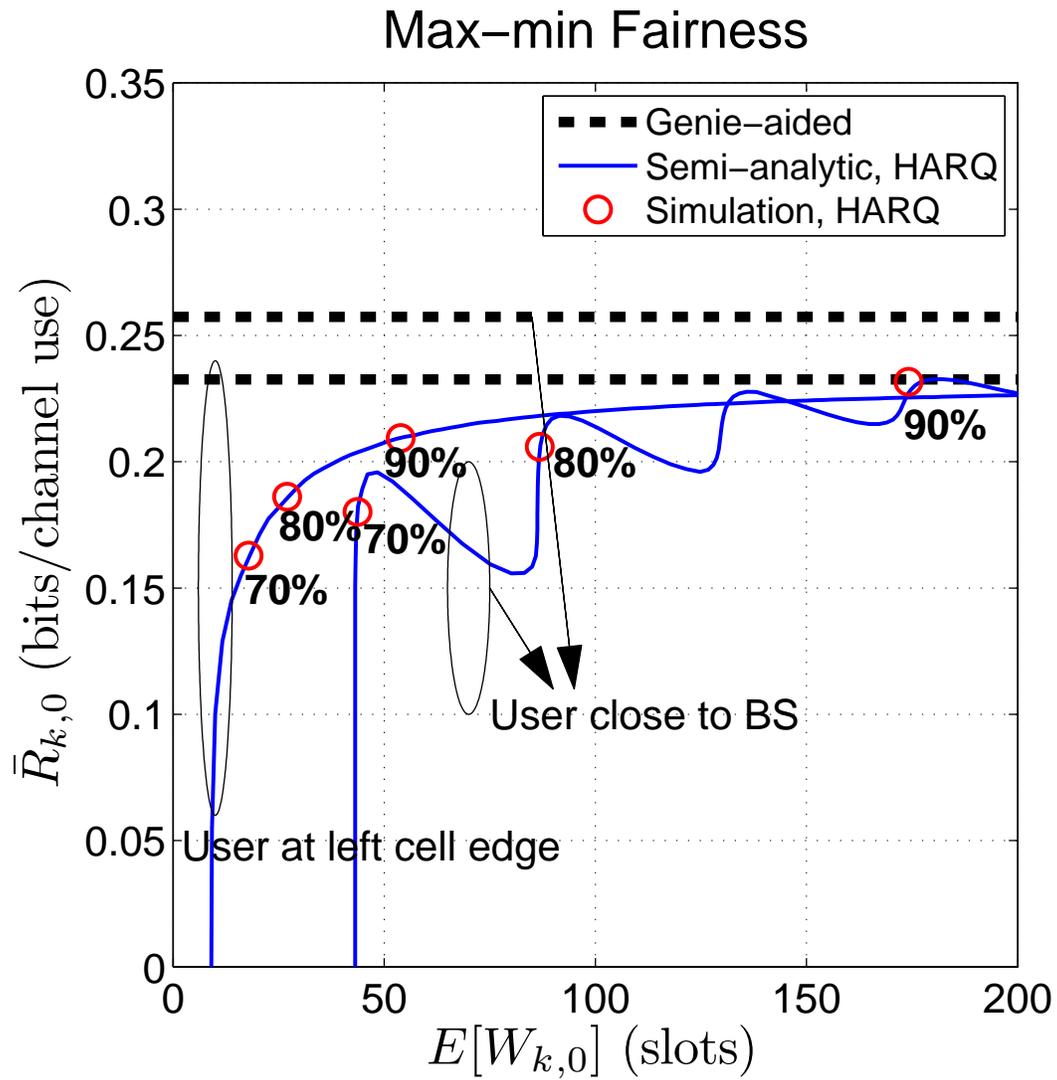}
  \caption{Average rate vs. decoding delay with max-min fairness  for two sample users.}
 \label{fig:rate_delay_HFS}
\end{figure}

Figs.~\ref{fig:rate_delay_PFS} and ~\ref{fig:rate_delay_HFS} illustrate the average throughput as a function
of the average decoding delay for the HARQ scheme in the case of two specific users: user
$(1, 0)$ at the left cell edge and $(18,0)$ at the cell center,
under proportional fairness and max-min fairness, respectively.
The thick dashed lines show genie-aided rates. The solid lines are obtained by increasing
first-block coding rate parameter $r_{k,0}$ and computing average decoding delay from (\ref{renewal})
with $\PP(\Ac_{k,0}[t])$ obtained by Monte Carlo simulation. Note that as $r_{k,0}$ increases,
also the delay $\EE[W_{k,0}]$ increases and the  HARQ throughputs approach the genie-aided
throughputs, in agreement with Theorem \ref{harq}. The ``o'' marks indicate the throughput-delay points at which the HARQ protocol
achieves 70\%, 80\% and 90\% of the genie-aided throughput based on simulations.
For example, under proportional fairness, 90\% of the genie-aided
throughput can be achieved at users $(1,0)$ and $(18,0)$ with average decoding delays of
about $57$ and $126$ slots, respectively. These points (obtained by full system simulation) are accurately predicted by the analytical formulas
of Section \ref{incr-harq} fitted with the Monte Carlo estimation of the probabilities $\PP(\Ac_{k,0}[t])$.

For $K = 36$ users per cell and $M = 2$ BS antennas, assuming that exactly $M = 2$ users are
served in each slot, a round-robin scheduling with no outage (genie-aided rate allocation)
would take an average delay of $18$ slots. Remarkably, under proportional fairness, 90\% of the genie-aided
throughput can be achieved with about $57$ slots of average delay for center user. This is only $\approx 3$ times
that of the genie-aided round-robin scheduling.
For edge users, this is achieved with $\approx 126$ slots of average delay for
the edge users, which is only $7$ times that of round-robin.
Under max-min fairness, both users $(1, 0)$ and $(18, 0)$ achieve genie-aided throughputs close to
$0.25$ bits/channel use. The decoding delay for the center user is larger than for the edge user
due to the fact  that center users are scheduled very rarely. For the $70\%$ point,
edge users achieve  $0.16$ bits/channel use with average delay of $18$ slots while center users
achieves a similar throughput of $0.18$ bits/channel use with delay of $44$ slots.

%

\section{Concluding remarks} \label{conclusions}

In this work we considered decentralized downlink scheduling in a multi-cell environment with multi-antenna BSs,
where the scheduler at each BS has perfect CSIT about its own users and {\em statistical} information about the
ICI caused by the other cells. Since each BS modifies its transmit covariance matrix at every slot,
the ICI powers experienced at the users' receivers are random variable.
We addressed the scheduling problem in the presence of uncertain ICI powers
in the framework of stochastic network optimization. A straightforward application of this framework yields
a conventional scheme based on adaptive variable-rate coding at the physical layer, and ARQ at the Logical Link Control layer.
Then, a new combination of the same stochastic network optimization framework with incremental redundancy Hybrid ARQ at the
physical layer was shown to improve over the conventional scheme, and achieve a network utility arbitrarily close to the performance
of a genie-aided system that can schedule the user rates equal to the (non-causally known) instantaneous mutual information on each slot.
For this scheme, we also showed that all Nash equilibria of the multi-cell decentralized scheduling game
yield network utility values that can be uniformly upper and lower bounded by virtual systems corresponding to ``extremal'' ICI statistics, where the
lower bound corresponds to the case of deterministic ICI powers equal to their mean values,
and  the upper bound corresponds to the case where all interfering BSs transmit to a single user
at full power (rank 1 interfering covariance matrices). These bounds stay at a fixed gap that is independent
of the cellular system configuration, i.e.,  of the channel path gain coefficients and operating SNR.
The proposed incremental redundancy HARQ can be implemented in practice by using Raptor codes \cite{RaptorCode} at the physical layer,
and need no protocol overhead to communicate slot-by-slot rate allocation as in adaptive variable-rate coding.
Hence, the proposed HARQ scheme is both easier to implement and performs significantly better than the conventional variable-rate coding scheme.
Also, we hasten to say that our approach applies directly to a variety of possible configurations,
including different MU-MIMO  precoding schemes and network MIMO schemes with clusters of coordinated cells \cite{Allerton08}.
In this paper we considered the case of linear ZFBF and no cell clustering for the sake of clarity of exposition.
The approach can also be extended to the case of non-perfect CSIT, following \cite{Hooman-nonperfectCSI}.
Here we focused on perfect CSIT for its simplicity and in order to focus on the random nature of ICI as the fundamental source of uncertainty
in a multi-cell environment.


\newpage

\bibliographystyle{IEEEtran}
\bibliography{biblio,2008_08_23_netmimo_v5,ETRI-report}

\newpage

\newpage

\newpage

\newpage

\newpage

%
%
%
%

\end{document}